\documentclass[preprints,article,accept,pdftex,moreauthors]{Definitions/mdpi} 
\firstpage{1} 
\pubvolume{1}
\issuenum{1}
\articlenumber{0}
\pubyear{2025}
\copyrightyear{2025}
\datereceived{ } 
\daterevised{ } 
\dateaccepted{ } 
\datepublished{ } 
\hreflink{https://doi.org/} 

\Title{Towards a Twisted Atom Laser: Cold Atoms Released from Helical Optical Tube Potentials}

\TitleCitation{Towards a Twisted Atom Laser: Cold Atoms Released from Helical Optical Tube Potentials}

\Author{Amine Jaouadi $^{1,*}$\orcidA{}, Andreas Lyras $^{2}$ and Vasileios E. Lembessis $^{2,*}$\orcidC{}}

\AuthorNames{Amine Jaouadi, Andreas Lyras, Vasileios E. Lembessis}

\isAPAStyle{%
       \AuthorCitation{Amine Jaouadi, Andreas Lyras, Vasileios E. Lembessis.}
         }{%
        \isChicagoStyle{%
        \AuthorCitation{Amine Jaouadi, Andreas Lyras, Vasileios E. Lembessis.}
        }{
        \AuthorCitation{Jaouadi, A.; Lyras, A.; Lembessis, V. E.}
        }
}

\address{%
$^{1}$ \quad LyRIDS, ECE Paris, Graduate School of Engineering, Paris 75015, France; ajaouadi@ece.fr\\
$^{2}$ \quad Photonics Group, Department of Physics and Astronomy, College of Science, King Saud
University, Riyadh 11451, Saudi Arabia}

\corres{Corresponding author: ajaouadi@ece.fr, vlempesis@ksu.edu.sa}

\abstract{
We study the quantum dynamics of cold atoms initially confined in a Helical Optical Tube (HOT) and subsequently released into free space. This helicoidal potential, engineered via structured light fields with orbital angular momentum, imposes a twisted geometry on the atomic ensemble during confinement. We examine how this geometry shapes the initial quantum state—particularly its spatial localization and phase structure—and how these features influence the subsequent free evolution.  Our analysis reveals that the overall confinement geometry supports the formation of spatially coherent, structured wavepackets, paving the way for the realization of twisted Bose–Einstein condensates and directed atom lasers. The results are of particular interest for applications in quantum technologies, such as coherent atom beam shaping, matter-wave interferometry, and guided transport of quantum matter.}


\begin{document}




\section{Introduction}

The cooling and trapping of atomic motion paved the way for a huge number of advances in physics with spectacular achievements in optical tweezing, atom optics, quantum simulations, quantum computing, materialization of gedankenexperiments, atom metrology and other \cite{cohen2011advances},\cite{meystre2001atom}, \cite{lembessis2020taming}. A major achievement was the first creation of a Bose-Einstein Condensate (BEC) from a dilute alkali atom gas by the research teams of C. Wieman and E. Cornell at JILA, Colorado,
and W. Ketterle at MIT. Both teams achieved Bose–Einstein condensation using sodium atoms in 1995 and these scientists were awarded the Nobel Prize in Physics in 2001 \cite{hess1986evaporative}, \cite{anderson1995observation},\cite{davis1995bose}, \cite{colombe2007strong}.

The transition of atoms from the disorder of the gaseous phase into the
“disciplined” behavior of the condensate can be compared with the
transformation of chaotic incoherent light into a laser light beam. This analogy
leads directly to the definition of the atom laser: a device that produces a beam
of boson atoms that have “escaped” from a trapped Bose–Einstein condensate \cite{robins2013atom}.
The atom laser plays a crucial role in advancing the field of atom optics, much like how the optical laser was instrumental in the swift progression of quantum optics fifty years ago \cite{meystre2001atom}. Atom laser plays important role in the development of research areas such as, quantum lithography, atom holography, quantum interference of matter waves, gravitational waves detection, non-linear atom optics, integrated atom optics and atomic solitons \cite{boto2000quantum},\cite{fujita2000interferometric}, \cite{becker2018space}, \cite{lenz1993nonlinear},\cite{salieres1999study}, \cite{khaykovich2002formation}. 

The arsenal of cold atom physics has been fortified since the advent of optical vortices \cite{allen1992orbital}. These are structured laser beams the photons of which carry an orbital angular momentum (OAM) apart from the spin angular momentum (SAM) which is associated with right and left circular polarization. The OAM carried by an optical vortex photon is $\ell\hbar$ with $\ell$ being an integer number called the winding number \cite{allen1999}. Optical vortices changed the cold atom physics landscape for two reasons \cite{babiker2018atoms}: a) they can impart an angular momentum to the atoms so they exert  torque on them and, thus, can rotate them around the beam propagation axis, and b) their cross-sectional intensity profile is characterized by azimuthally symmetric bright and dark regions leading to the formation of cylindrically symmetric optical dipole traps and optical lattices like the optical Ferris wheel, which are formed by the superposition of two similar co-propagating Laguerre-Gaussian (LG) beams with opposite winding numbers and they are characterized by a transverse pedal-like intensity pattern \cite{franke2007optical}, \cite{lembessis2020optical}. The idea of the optical Ferris wheel has been also extended to cold atom beams with a transverse probability density of similar structure \cite{lembessis2017atomic}, \cite{lembessis2018light}.

A three-dimensional version of a Ferris wheel is the so called helical optical tubes (HOT) which are formed by the interference of two counter-propagating LG beams with opposite winding numbers. Helical Optical Tube (HOT) potentials have emerged as a groundbreaking platform for trapping and manipulating quantum particles, particularly ultracold atoms and Bose-Einstein condensates (BECs) \cite{okulov2013superfluid}, \cite{okulov2020self}, \cite{lembessis2022quantum}, \cite{al2016rotating}, \cite{al2016guiding}. These potentials are created by the interference of counter-propagating Laguerre-Gaussian laser beams with opposite winding numbers. Their intricate helical structure enables unique trapping potentials and control over quantum states. The study of cold atoms deeply trapped by the optical dipole potential of a HOT has shown that the spatial atom wavefunctions have a helical structure \cite{lembessis2022quantum}. 

This intriguing feature brings to the fore the question of the form of the probability density which the atom would have once it is released from such a trap and propagates into free space.  The time-dependent Schrödinger and/or Gross-Pitaevskii equations are solved using the split-step Fourier method, alternating between position space for potential and interaction terms and momentum space for kinetic terms \cite{Kosloff1986}. The visualization is performed using the Mayavi package \cite{Mayavi}. Our numerical analysis shows that the atom or even a BEC released from such a trap are characterized by probability densities which are twisted in space. This property could be exploited for the generation of novel types of atom laser beams where the atoms will follow a helical path while propagating in free-space. Our findings highlight the localized nature of the wavefunction under gravitational expansion. Remarkably, the wavefunction remains strongly confined. This spatial localization underscores the HOT potential's capacity to maintain coherence and precise control over quantum states, even in dynamically evolving conditions. Spatial localization is crucial for maintaining quantum coherence, enabling precise control, and facilitating various applications in quantum technologies. Localization minimizes decoherence, enhances measurement precision, and allows controlled manipulation of quantum systems. This is essential in fields like quantum computing, atom interferometry, ultracold physics, and quantum sensing. This new family of atom beams will expand the scope of atom optics, as the beams will exhibit sensitivity to rotations and structurally richer interference patterns. In addition to linear momentum, their angular momentum and helicity will serve as crucial parameters, further enriching their potential applications and insights. 

\section{Materials and Methods}

\subsection{Optical Potential in Helical Coordinates}


We begin with the formulation of the optical dipole potential generated by a \textit{Ferris wheel}--type light field, which arises from the interference of two counter-propagating Laguerre-Gaussian (LG) beams with opposite helicities, i.e., winding numbers $\pm\ell$ where $\ell > 0$. In the far-detuned limit, the time-averaged dipole potential experienced by a neutral atom is given by:
\begin{equation}
U(\rho,\phi,z) = V_0 \left( \frac{\sqrt{2}\rho}{w(z)} \right)^{2|\ell|} \exp\left(-\frac{2\rho^2}{w(z)^2}\right)\cos^2(kz + \ell\phi),
\label{eq:potential}
\end{equation}
where $w(z) = w_0\sqrt{1 + (z/z_R)^2}$ is the $z$-dependent beam waist, $k$ is the optical wave number, and $V_0$ is the trap depth determined by the beam intensity and detuning. This potential features both radial and azimuthal confinement, while the term $\cos^2(kz + \ell\phi)$ introduces a helicoidal modulation along the beam axis.

\begin{figure}[h!]
    \centering
    \includegraphics[width=0.5\textwidth]{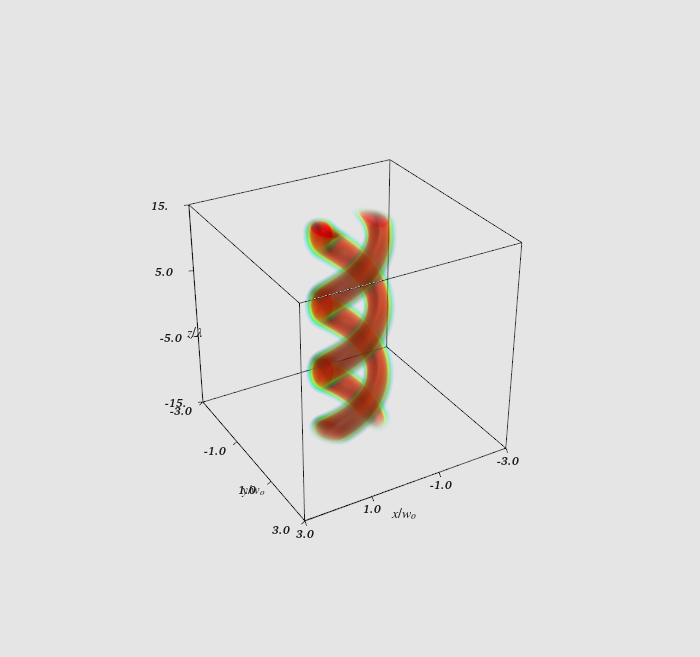}
    \caption{Three-dimensional visualization of the HOT (Helical Optical Trap) potential for winding number $\ell = 1$. The potential forms a single helix of local minima that guides atomic motion along a twisted path.}
    \label{fig:HOT_potential}
\end{figure}

The resulting landscape forms a three-dimensional helical lattice structure, with $\ell$ intertwined potential minima winding around the optical axis. This structure serves as a waveguide for atomic matter waves and is particularly suitable for producing coherent guided atomic beams. The spatial profile of the potential for the case $\ell = 1$ is visualized in Figure~\ref{fig:HOT_potential}, illustrating the helicoidal geometry and the confinement mechanism.

To exploit the helical symmetry of the optical potential, we briefly recall the coordinate transformation into Fresnel coordinates, which align naturally with the helicoidal geometry induced by interfering Laguerre-Gaussian beams of opposite winding numbers. These coordinates $(\rho, \nu, \xi)$, introduced in our earlier work \cite{lembessis2021}, allow simplification of the dipole potential into a longitudinally modulated form:
\begin{equation}
    U(\rho, \xi, \nu) = \frac{\hbar \Omega_0^2}{\Delta} \, u_{|l|}^2(\rho, \xi) \cos^2(k \nu),
    \label{eq:helicoidal_potential_short}
\end{equation}
where $u_{|l|}^2(\rho, \xi)$ encapsulates the radial Gaussian envelope centered around the intensity maximum. In the paraxial regime ($z \ll z_R$) and under weak radial deviations---meaning small transverse displacements from the beam axis such that $\rho^2 \ll w_0^2$ - atoms remain tightly confined near the helical troughs of the potential. The depth of this structure is quantified by the characteristic energy scale:
\begin{equation}
    \varepsilon = \frac{\hbar \Omega_0^2}{\Delta} \, \frac{|l|^{|l|}}{|l|!} e^{-|l|},
    \label{eq:depth_compact}
\end{equation}
which governs the localization and dynamics of atoms within the twisted guide.

\subsection{Stationary Schrödinger Equation and Ground State}
\subsubsection{Two-Level Atom }
The stationary states of a single atom confined in the helical optical potential of Eq.~\eqref{eq:potential} have been analytically derived in previous work~\cite{lembessis2022quantum}. By transforming to Fresnel coordinates aligned with the helical geometry, the wavefunction separates as:
\begin{equation}
    \psi_{n_\rho,n_\nu,n_\xi}(r, \phi, z) = R_{n_\rho}(\rho) \, N_{n_\nu}(\nu) \, \Xi_{n_\xi}(\xi),
\end{equation}
where $\rho = r - \sqrt{|l|/2}\,w(z)$ is the radial displacement from the intensity ring, $\nu = z + l\phi/k$ encodes the helical phase, and $\xi = z/h$ is the longitudinal coordinate. The functions $R_{n_\rho}$ and $N_{n_\nu}$ are harmonic oscillator eigenstates, while $\Xi_{n_\xi}$ satisfies a Mathieu-type equation.

The energy spectrum is given by:
\begin{equation}
    E_{n_\rho,n_\nu,n_\xi} = -\epsilon + \hbar \omega_\rho \left(n_\rho + \tfrac{1}{2} \right) + \hbar \omega_\nu \left(n_\nu + \tfrac{1}{2} \right) + \Delta E(n_\xi),
\end{equation}
where the confinement frequencies $\omega_\rho$, $\omega_\nu$ and potential depth $\epsilon$ depend on laser parameters and the winding number $\ell$. 

\begin{figure}[h!]
\centering
\includegraphics[width=0.5\textwidth]{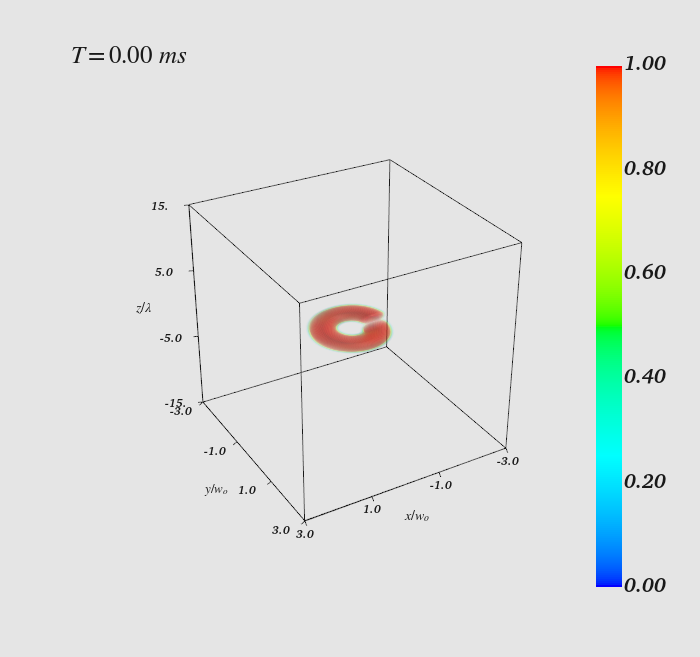}
\caption{3D plot of the ground-state probability distribution $|\psi_{0,0,0}(x,y,z)|^2$ with experimental parameters: $P = 5$~mW, $\Delta = -10^{15}$~Hz, $w_0 = 4~\mu$m, $l = 1$, $\epsilon = 41.6 E_r$.}
\label{fig:ground_state}
\end{figure}

The ground state wavefunction exhibits a helical density distribution centered around the optical intensity ring (Fig.~\ref{fig:ground_state}), serving as the initial state for free-fall dynamics presented in the next sections. The case taken here is for a single Rb atom trapped by a HOT field which interacts with the transition $5^2S_{1/2}-5^2P_{3/2}$ with a total power $P=5$  mW, a detuning $\Delta=-10^{15}$ Hz a beam waist $w_{0}=5\ \mu$m, winding numbers $\ell=\pm1$ and $\epsilon=41.6 \ E_{r}$ with $E_{r}$ being the recoil energy associated with the Rb transition.

\subsubsection{Bose-Einstein Condensate }

In the Thomas-Fermi (TF) regime, where the kinetic energy is negligible compared to the interaction energy between the $N$ atoms composing the BEC, the wavefunction of the Bose--Einstein condensate (BEC) is determined by the balance between the trapping potential \( U(\rho, \varphi, z) \) and the repulsive interactions. The TF wavefunction is given by:
\begin{equation}
    \psi(\rho, \varphi, z) = 
    \begin{cases}
        \sqrt{\dfrac{\mu - U(\rho, \varphi, z)}{Ng_{int}}}, & \text{for } U(\rho, \varphi, z) < \mu, \\
        0 & \text{otherwise},
    \end{cases}
    \label{eq:tf_wavefunction}
\end{equation}
where \( \mu \) is the chemical potential, determined by normalizing the wavefunction to the total number of atoms \( N \), and \( g_{int} = \dfrac{4\pi \hbar^2 a_s}{m} \) is the interaction strength, with \( a_s \) denoting the \( s \)-wave scattering length and \( m \) the atomic mass. In our simulations, we consider \( ^{87}\mathrm{Rb} \) atoms under typical experimental conditions.
\( U(\rho, \varphi, z) \) is the optical dipole potential (Eq.~(2)), which depends on the angular \( \varphi \), radial \( \rho \), and axial \( z \) coordinates.

The BEC occupies the region where the external potential satisfies \( U(\mathbf{r}) < \mu \), resulting in a Thomas--Fermi density profile that inversely reflects the shape of the trapping potential. Notably, while the overall density scales with the number of atoms \( N \), the shape of the Thomas--Fermi distribution remains unchanged, as it is governed solely by the geometry of the potential and the balance of interactions in the Thomas--Fermi regime.

\section{Free Fall Under Gravity}

Once the HOT is turned off, the atom experiences free fall governed by the time-dependent Schrödinger equation:
\begin{equation}
i\hbar\frac{\partial \Psi}{\partial t} = \left[-\frac{\hbar^2}{2m}\nabla^2 + mgz \right]\Psi.
\end{equation}

We simplify this equation by transforming to a freely falling reference frame \cite{Gaaloul2007, Gaaloul2009}:
\begin{equation}
\Psi(\mathbf{r},t) = \exp\left[\frac{im}{\hbar}g t z + \frac{img^2 t^3}{6\hbar} \right] \Xi(\mathbf{R},t),
\end{equation}
with $\mathbf{R} = \mathbf{r} + \frac{1}{2}gt^2\hat{z}$, leading to:
\begin{equation}
i\hbar \frac{\partial \Xi}{\partial t} = -\frac{\hbar^2}{2m} \nabla^2 \Xi.
\end{equation}

In this new equation, the gravitational term \( mgz \) no longer appears. The macroscopic wavefunction \( \Xi(\mathbf{R}, t) \) now describes the dynamics in a free-falling reference frame, greatly simplifying numerical simulations and analytical treatments. This transformed equation describes a free evolution with an initial condition given by the wavefunction in the HOT potential. 

The time-dependent Schrödinger equation and, where relevant, the Gross--Pitaevskii equation are numerically integrated using the split-step Fourier method \cite{Kosloff1986}. This approach efficiently handles the nonlinear and kinetic contributions by alternating between position and momentum space at each time step: the nonlinear interaction and potential terms are evaluated in real space, while the kinetic evolution is computed in Fourier space. This method ensures stable and accurate propagation of the wavefunction over extended simulation times.

For our simulations, we consider $^{87}$Rb atoms as the working species, a standard choice in cold atom experiments due to their well-characterized properties and wide experimental relevance. The optical trapping potential is generated by a structured laser beam with power $P_r = 5$~mW, detuning $\Delta = -100 \times 10^{13}$~Hz (red-detuned), and beam waist $w_0 \approx 4~\mu$m. We set the orbital angular momentum quantum number to $l=1$, corresponding to a first-order Laguerre--Gaussian beam. The trap depth is fixed at $\varepsilon = 41.6~E_r$, where $E_r$ denotes the recoil energy of the atom.

\section{Results and Discussion}

\subsection{Two-Level Atom}

\begin{figure}[h!]
\centering
\includegraphics[width = 1\textwidth]{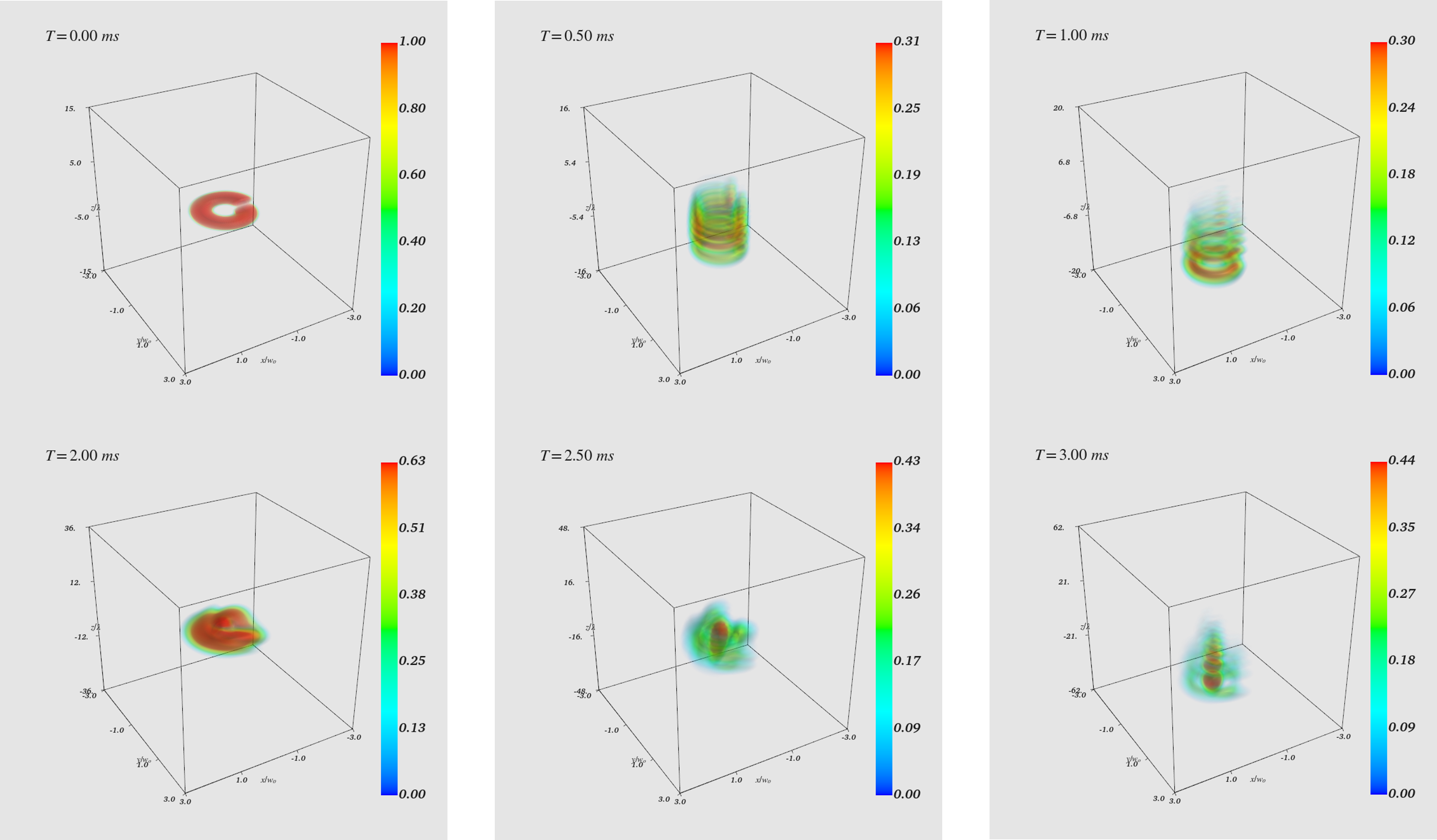}
\caption{3D plot of the ground-state probability distribution $|\psi_{0,0,0}(x,y,z)|^2$ and its evolution under gravity. The experimental parameters: $P = 35$~J/s, $\Delta = -10^{15}$~Hz, $w_0 = 4~\mu$m, $l = 1$, $\epsilon = 41.6 E_r$.}
\label{fig:ground_state2}
\end{figure}

Upon release from the helicoidal optical trap (HOT), the initially ring-shaped atomic wavepacket—previously shown in Fig.~\ref{fig:ground_state2}—evolves solely under the influence of gravity. With the trapping potential turned off, the external field reduces to a linear gravitational potential that induces a uniform downward acceleration of the wavepacket's center of mass at a rate \( g \).

At early times (\(T = 0.00\)–\(0.50\) ms), the wavepacket exhibits rapid radial expansion due to kinetic dispersion. This results in a reduction of peak probability density—a phenomenon we refer to as \textit{dilution}, stemming from the spread of the wavefunction into a larger spatial volume rather than any dissipative process.

Between \(T = 0.50\) ms and \(T = 1.00\) ms, the wavepacket begins to develop concentric interference fringes. These fringes are a direct result of self-interference among different portions of the expanding ring-shaped wavefunction, which maintain a high degree of phase coherence from the initial state. The cylindrical symmetry of the initial condition favors the emergence of these toroidal interference features.

At \(T = 2.00\) ms, the density distribution undergoes a noticeable temporary "refocusing," where the peak density briefly increases before continuing its overall decay. This resurgence may be attributed to constructive self-interference enabled by the initial ring geometry, which imposes a curved wavefront topology even during free fall.

From \(T = 2.50\) ms onward, the expansion continues and the interference pattern gradually loses contrast. By \(T = 3.00\) ms, the wavepacket has spread significantly, and the peak density settles to approximately 40–45\% of its original maximum. The observed spatiotemporal evolution highlights the interplay between coherence, geometry, and dispersion in freely expanding wavefunction.

The revival dynamics is a result of the geometry of the trap, which combines longitudinal motion (along the helix) with azimuthal/angular motion. Upon release, this angular momentum is converted into a structured free-space evolution, akin to a rotating and shearing wave packet. This leads to self-interference of the wavefunction in space-time, which, under the right conditions, causes refocusing. Once we release the atom from the trap, the time evolution of its wavefunction is determined by the quantities $T_{\rho}=2\pi/\omega_{\rho}$, $T_{\nu}=2\pi/\omega_{\nu}$ and an evolution time corresponding to the $\xi$ coordinate. The latter is given by $T_{\xi}=2\pi\hbar/\Delta E$, where $\Delta E=2\alpha^{3/4}\ell\sqrt{\epsilon E_{r}}$ is the energy gap between two adjacent states, with $E_{r}$ being the recoil energy. Straightforward calculations for the parameter values used give us $T_{\rho}\approx 7.46$ ms, $T_{\nu}\approx 0.062$ ms, and $T_{\xi}\approx 2$ ms. The numerical work clearly shows that the factor that drives revivals is the time evolution due to the variable $\xi$. At first sight this may seem not expected since the slowest variable is $T_{\rho}$, which means that either radial excitations are nearly frozen (i.e. $n_{\rho}=0$, or contribute broad, slowly varying envelopes but not the sharp revival feature. This is consistent with tight radial trapping as in the case that has been considered in \cite{al2016guiding}. The coordinate $\xi$ describes motion along the helicoidal trough of the potential, which is periodically modulated. This modulation leads to a Mathieu equation, giving a spectrum which corresponds to a quasi-harmonic ladder. So the Mathieu mode creates the internal clock that governs revival dynamics in our system.

These observations have several important implications. First, a clear demonstration of free-fall dynamics and refocusing validates the use of ring-shaped wavefunctions as sensitive probes. Second, dispersive fringe patterns may be exploited for matter wave interferometry, since their spatial period encodes the local phase gradient and hence any additional accelerations or potentials. Finally, refocusing provides a means to transiently enhance density - and thus nonlinearity - without reactivating the trap.

Extending these experiments to microgravity environments could enable even longer interrogation times and more pronounced phenomena, with potential applications in precision metrology.

\subsubsection{Influence of the beam waist}

\begin{figure}[htbp]
    \centering
    \includegraphics[width=\textwidth]{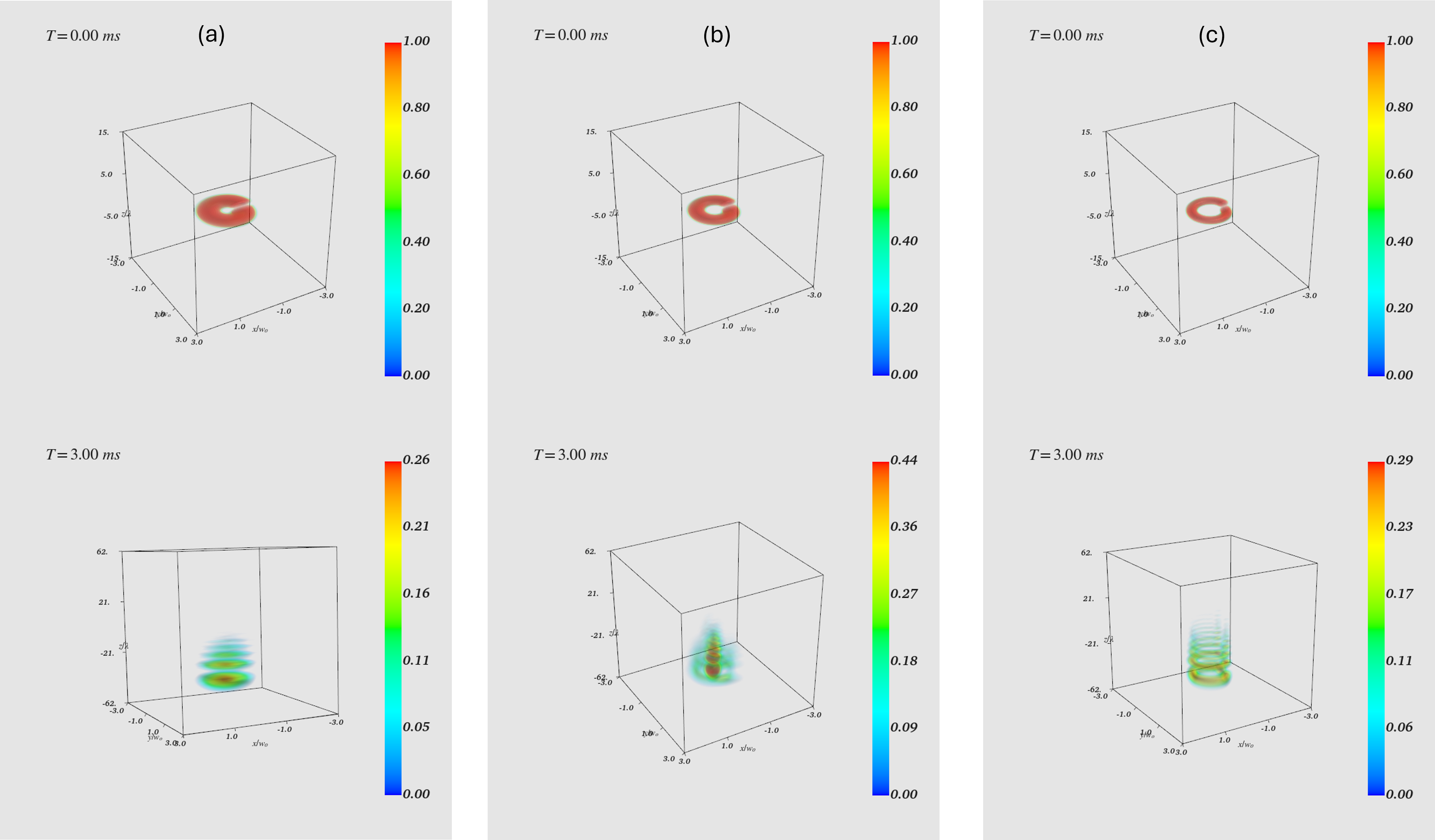}
    \caption{Time evolution of the wavefunction of a single atom confined in a helicoidal optical trap (HOT) and released under gravity. Each column corresponds to a different beam waist \( w_0 \), increasing from left to right. Panel (a) for \( w_0 = 2\  \mu m \), panel (b) for \( w_0 = 4\  \mu m \) and panel (c) for \( w_0 = 10\  \mu m \). The top row shows the initial probability density at \( T = 0.00 \, \text{ms} \), while the bottom row shows the evolved density after \( T = 3.00 \, \text{ms} \) of free fall. Colorbars represent normalized probability density.}
    \label{fig:1atom_waist}
\end{figure}

Figure~\ref{fig:1atom_waist} illustrates the evolution of the probability density of a single atom initially confined to a HOT and subsequently undergoing a free fall under gravity. The simulation is performed for different values of the optical beam waist \( w_0 \), which determines the radial confinement of the trap. The atom is initially prepared in the ground state of the HOT potential.

At \( T = 0 \, \text{ms} \), the probability density is localized in a ring-shaped region in the \(( x\text{-}y )\) plane, with strong confinement in the \( z \)-direction. As expected, smaller beam waists lead to tighter radial confinement and more compact toroidal shapes. As \( w_0 \) increases, the radius of the ring and the radial width also increase, reflecting the larger spatial extent of the potential minimum.

After \( T = 3.00 \, \text{ms} \) of evolution under gravity, the wavefunction exhibits downward motion due to gravitational acceleration. The final probability densities reveal significant vertical spreading and deformation, with features strongly dependent on the initial radial confinement. For smaller beam waists (panel (a)), the wavefunction spreads rapidly and loses its ring-like structure, indicating strong dispersion. In contrast, for larger \( w_0 \) (panel (c)), the atom retains more structural coherence, and periodic features emerge in the vertical direction. These modulations result from the interplay between the initial spatial confinement, gravitational acceleration, and the geometry of the helical optical tube, which influences the redistribution of density during free fall.

\subsubsection{Dynamics for longer time}

 We analyze the free-fall dynamics of a single atom initially prepared in the ground state of a Helical Optical Tube (HOT) with a beam waist \( w_0 = 4\, \mu\mathrm{m} \) for a larger evolution time \( T = 5 \ ms \). The time evolution of the wavefunction density exhibits clear evidence of periodic refocusing, which reflects the interplay between the initial spatial confinement and the dispersive dynamics under gravity. Below are some key observations:
 
At \(T = 0\, \mathrm{ms}\), the wavefunction shows a well-defined ring-shaped density, consistent with the trapped ground state in the HOT.

Between \(T \sim 0.25 - 1.75\, \mathrm{ms}\) the wavefunction undergoes rapid radial expansion and axial stretching due to free propagation and gravity. Interference fringes emerge, revealing phase evolution across the expanding toroidal wavefront.

At \(T \sim 2.25\, \mathrm{ms}\) the wavefunction exhibits a first refocusing, with the reappearance of the ring-like structure and central localization.

During \(T \sim 2.50 - 3.75\, \mathrm{ms}\), the density becomes axially elongated and less structured, suggesting gravitational stretching and partial decoherence.

At \(T \sim 4.25\, \mathrm{ms}\), a second refocusing is observed, with stronger central localization and a reconstructed toroidal shape.

After \(T > 4.25\, \mathrm{ms}\), the density again loses its ring-like coherence as the wavefunction continues to stretch.

This behavior indicates a quasiperiodic refocusing mechanism, where the initial ring-shaped structure is partially restored at regular time intervals. The reappearance of similar shapes around \(T \sim 2.25\, \mathrm{ms}\) and \(T \sim 4.25\, \mathrm{ms}\) suggests a characteristic refocusing time scale:
\[ T_f \sim 2\, \mathrm{ms}.\]

Interestingly, refocusing times appear to be independent of initial confinement strength, particularly the beam waist \(w_0\). For example, the wavefunction regains its original ring-shaped profile at similar times (\(T \sim 2.25\,\mathrm{ms}\) and \(T \sim 4.25\,\mathrm{ms}\)) for larger waists such as \(w_0 = 6\ \mu\mathrm{m}\) and \(w_0 = 10\ \mu\mathrm{m}\), as it does for \(w_0 = 4\ \mu\mathrm{m}\). This suggests that the periodicity of refocusing is primarily governed by the interplay between kinetic and gravitational energy scales, rather than by the details of the initial spatial confinement. 
In contrast, for tighter confinement (\(w_0 = 2\ \mu\mathrm{m}\)), while refocusing still occurs, the wavefunction fails to fully retrieve its initial ring shape. Instead, it exhibits a more spheroidal form as a result of stronger initial localization and enhanced phase dispersion. At later times (\(T > 4.5\,\mathrm{ms}\)), significant spatial spreading and amplitude decay are observed, indicative of gravitational spreading.

\subsection{ BEC}


\subsubsection{$\ell =1$}

\begin{figure}[H]
\centering
\subfloat[\centering]{\includegraphics[width=6.0cm]{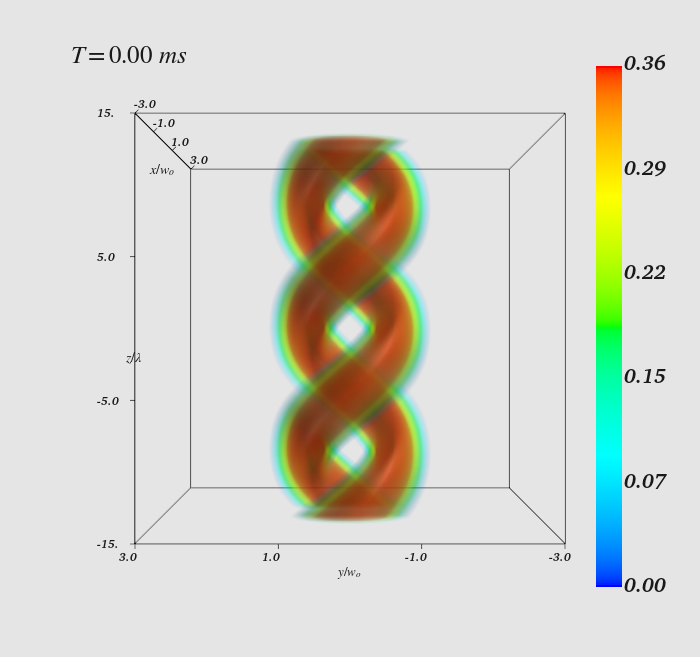}}
\subfloat[\centering]{\includegraphics[width=6.0cm]{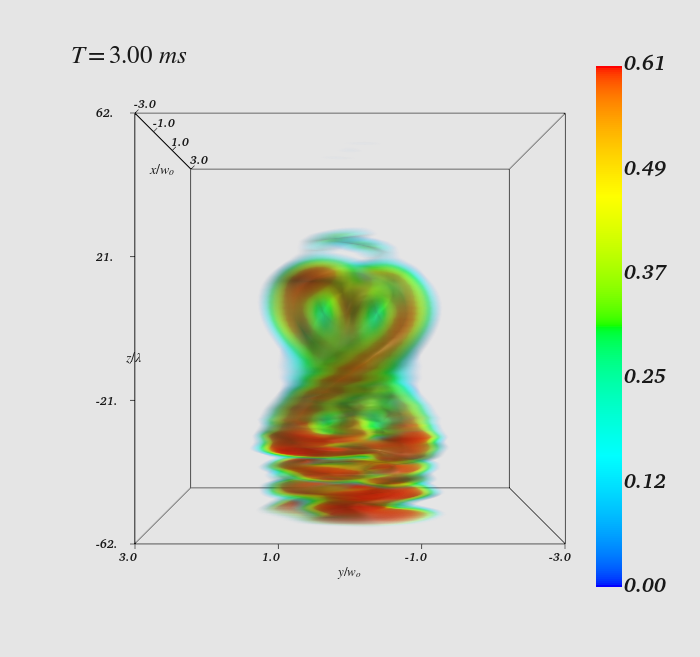}}
\hfill
\subfloat[\centering]{\includegraphics[width=6.0cm]{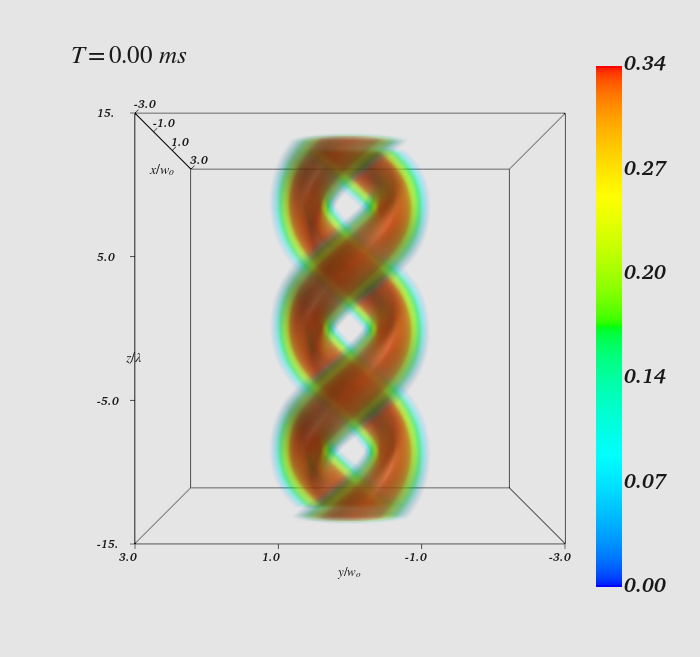}}
\subfloat[\centering]{\includegraphics[width=6.0cm]{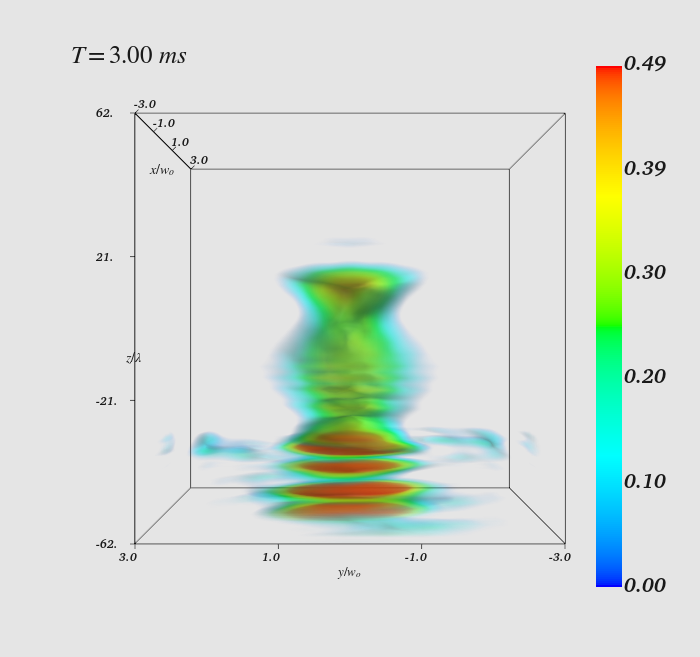}}
\caption{Time evolution of a BEC in a helicoidal optical trap (HOT) under gravity. Top row: beam waist \( w_0 = 4\,\mu\text{m} \) at \( T = 0\,\text{ms} \) (panel (a)) and \( T = 3\,\text{ms} \) (panel (b)). Bottom row: same evolution for \( w_0 = 2\,\mu\text{m} \), panel (c) for \( T = 0\,\text{ms} \) and panel (d) for \( T = 3\,\text{ms} \). The color scale represents the density \( |\psi(\mathbf{r},t)|^2 \).}
\label{BEC_l=1_waist}
\end{figure}

Figure~\ref{BEC_l=1_waist} compares the three-dimensional evolution of a BEC confined in a helicoidal optical trap (HOT) and subjected to gravitational free fall, for two different beam waists: \( w_0 = 4\,\mu\text{m} \) (top row) and \( w_0 = 2\,\mu\text{m} \) (bottom row). One should recall that the parameter \( \ell = 1 \) sets the helicoidal pitch, while the beam waist controls the radial confinement and overall intensity profile of the trap.

At initial time \( T = 0 \,\text{ms} \), both configurations exhibit a pronounced helicoidal density pattern along the \( z \)-axis, characteristic of the HOT-induced potential landscape. The helicity is more sharply defined for the narrower beam waist (\( w_0 = 2\,\mu\text{m} \)) (panel (c)), indicating tighter radial localization.

After \( T = 3\,\text{ms} \) of free fall, significant differences in the dynamics emerge:

For \( w_0 = 4\,\mu\text{m} \) (panel (b)), the wavefunction spreads more uniformly during its descent. The helical structure partially washes out, but the condensate maintains an overall elongated columnar shape. This evolution is largely governed by gravitational acceleration and relatively weaker transverse confinement, which allows broader dispersion.

For \( w_0 = 2\,\mu\text{m} \) (panel (d)), the condensate exhibits nontrivial focusing behavior, forming a pronounced density peak around the trap axis during evolution. This self-focusing effect suggests enhanced nonlinear interactions due to initial higher density and tighter confinement, which act against the gravitational spreading and lead to compression in the vertical (\( z \)) direction. Moreover, the outer regions show signs of radial expansion and modulated interference rings, consistent with interaction-driven reshaping of the wavepacket.

Again this observation emphasizes the critical role of the beam waist in shaping both the initial wavefunction morphology and its subsequent nonlinear evolution. Tighter waists induce stronger radial localization, higher initial peak densities, and more prominent nonlinear effects, ultimately leading to phenomena such as transient self-focusing during gravitational descent. These features could be harnessed in controlled matter-wave shaping or soliton-based transport schemes within engineered optical landscapes \cite{Jaouadi2024, Jaouadi2025}.

\subsubsection{$\ell > 1$}

\begin{figure}[H]
\centering

\subfloat[\centering]{\includegraphics[width=6.0cm]{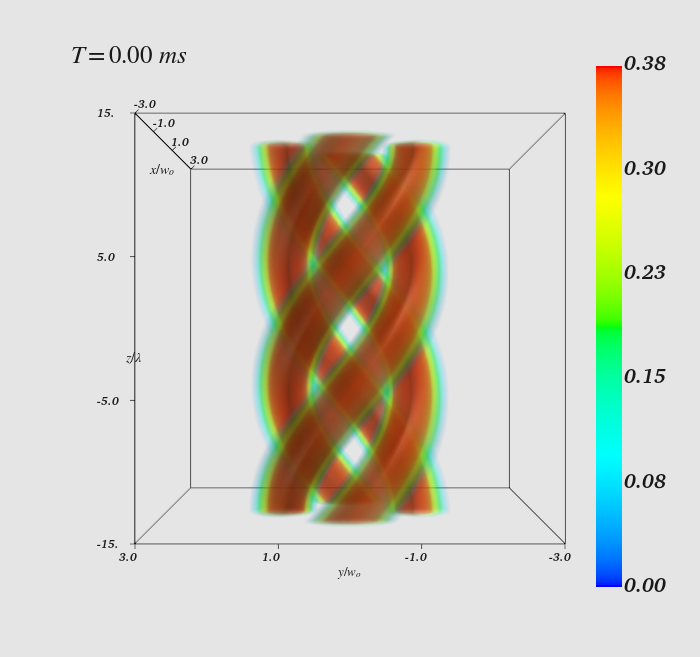}}
\subfloat[\centering]{\includegraphics[width=6.0cm]{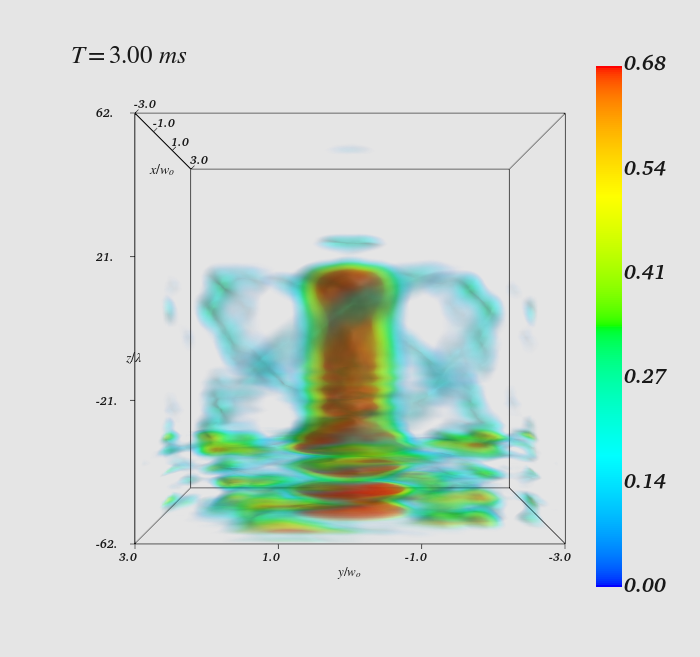}}
\hfill
\subfloat[\centering]{\includegraphics[width=6.0cm]{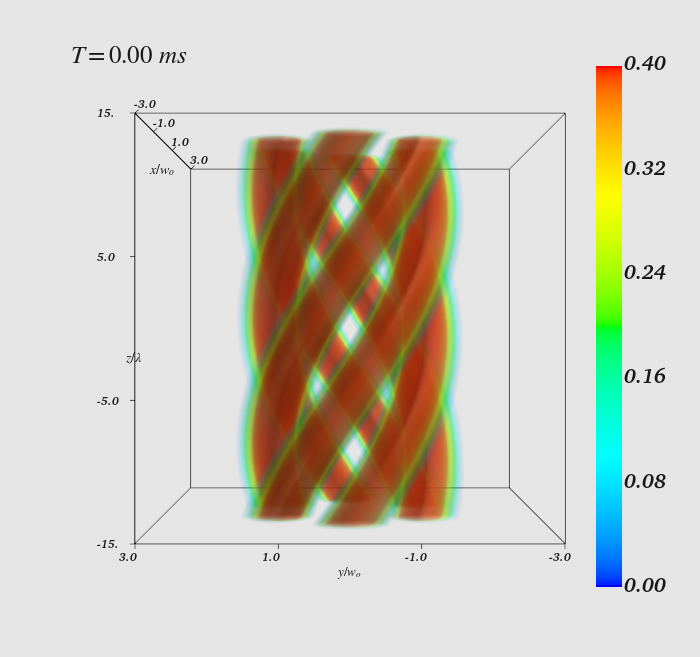}}
\subfloat[\centering]{\includegraphics[width=6.0cm]{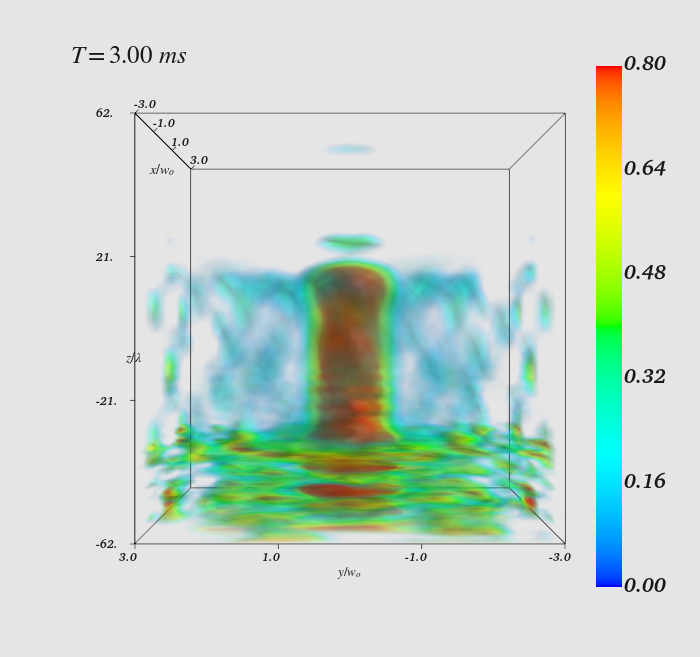}}
\caption{BEC evolution in a helicoidal optical trap with increased winding number. First column: \( \ell = 2 \), Second column: \( \ell = 3 \). Top row: initial states at \( T = 0\,\text{ms} \). Bottom row: evolved states at \( T = 3\,\text{ms} \). The colorbars indicate density values \( |\psi|^2 \).)
\label{BEC_l2_l3}}
\end{figure}

Figure~\ref{BEC_l2_l3} displays the evolution of a BEC confined in a HOT with higher winding numbers \( \ell = 2 \) (top row) and \( \ell = 3 \) (bottom row). The parameter \( \ell \) characterizes the number of turns per unit vertical length in the helicoidal potential, directly controlling the rotational modulation and twisting of the condensate along the \( z \)-axis. As previously, these simulations are performed under free fall for  \( T = 3\,\text{ms} \), enabling the observation of gravitational and nonlinear effects in the absence of confinement.

At \( T = 0 \,\text{ms} \), both configurations show highly structured density patterns with multiple intertwined strands reflecting the increased helicity. As \( \ell \) increases, the number of visible braids or twists in the condensate increases accordingly. For \( \ell = 3 \), the initial state exhibits a more tightly wound and spatially compressed structure, with 4 braids (panel (a)), than in the \( \ell = 2 \) case, which has 6 braids (panel (c)). 

At \( T = 3 \,\text{ms} \), after the trapping potential is turned off and the condensate falls under gravity, clear differences emerge in the density profiles:

- For \( \ell = 2 \): The BEC expands and forms a vertically stretched, column-like structure, with concentric rings and interference fringes appearing toward the lower region. The wavefunction remains relatively coherent, although interactions generate radial modulations indicative of nonlinear phase evolution.

- For \( \ell = 3 \): A pronounced central peak develops during free fall, with a sharper and more confined density maximum at the center of mass. This suggests a stronger self-focusing effect, likely driven by the more intricate initial twist structure and associated higher spatial confinement. The condensate appears to resist radial dispersion more effectively than in the \( \ell = 2 \) case where the read regions are close to 0.80. Simultaneously, strong interference fringes appear, especially in the lower region, indicating intensified nonlinear wave mixing during expansion.

This analysis reveals that increasing the helicoidal winding number \( \ell \) not only enriches the internal structure of the BEC in the trapped state but also significantly alters its dynamical evolution. Higher \( \ell \) values lead to stronger initial localization and more pronounced nonlinear effects during gravitational expansion. These results highlight the potential of HOTs with variable \( \ell \) as versatile platforms to control the spatiotemporal behavior of quantum matter waves, enabling precise tuning, focusing, and transport phenomena in ultracold atom experiments.

The spatio-temporal evolution of ultracold matter released from helicoidal optical traps (HOTs) under gravity reveals rich dynamical behavior, as illustrated through numerical simulations for both single atoms and interacting Bose--Einstein condensates (BECs). These dynamics provide insight into the design and control of atom-laser-like sources, particularly when structured phase and amplitude profiles are involved.

The single-atom case highlights the potential for coherent beam shaping without interaction-induced decoherence, while the BEC case offers rich nonlinear phenomena for advanced manipulation.

The observed self-focusing, phase imprinting, and helical unwrapping suggest that HOT-based architectures can generate twisted atom lasers—coherent matter-wave analogues of optical vortex beams. Such devices could be foundational for precision inertial sensing and interferometry with topological modes, guided matter-wave transport with angular momentum encoding, and quantum information protocols using orbital angular momentum states,

The gravitational release of both single atoms and condensates from HOTs opens a pathway to structured atom laser generation. The helicity parameter \( \ell \) and beam waist \( w_0 \) serve as key control knobs for tuning the beam shape, focus, and coherence, bridging nonlinear wave dynamics and quantum technology applications.

\section{Conclusions}

We have presented an analytical and numerical analysis of a trapped atom and a BEC in a Helical Optical Tube under gravity. Our work shows that the helical trap supports highly localized quantum states and that free fall dynamics are significantly affected by the geometry of the initial potential. 

The numerical simulations presented in this work shed light on the intricate dynamics of Bose–Einstein condensates released from helicoidal optical traps under gravity. By varying key trap parameters—such as the beam waist and the helicoidal winding number \(\ell\)—we have revealed a rich spectrum of free-fall behavior, including nonlinear focusing, helicoidal unwinding, and strong matter-wave interference. These effects stem from the interplay between initial helical confinement, gravitational acceleration, and interatomic interactions. In particular, tighter radial confinement and higher helicity were found to enhance nonlinear self-focusing and axial density redistribution, forming narrow, high-density output channels reminiscent of guided atomic beams.

These findings point toward a promising application: the realization of a wisted Atom Laser. By engineering the initial trap geometry, especially through helicoidal phase imprinting and tunable winding, it becomes feasible to control the directionality, coherence, and angular momentum of atom-laser output. Such a twisted matter-wave source could be of significant interest for precision sensing, inertial navigation, and probing rotational quantum phenomena. The ability to shape and manipulate atomic beams using optical topology marks an important step toward atom-optical devices with tailored spatial modes and controlled output geometries.


\vspace{6pt} 





\authorcontributions{Conceptualization, A.J., A.L. and V.E.L.; methodology, A.J., A.L. and V.E.L.; software, A.J.; validation, A.J., A.L. and V.E.L.; formal analysis, X.X.; investigation, A.J., A.L. and V.E.L.; writing---original draft preparation, A.J. and V.E.L.; writing---review and editing, A.J., A.L. and V.E.L.; visualization, A.J. All authors have read and agreed to the published version of the manuscript.}

\funding{Please add: This research received no external funding}

\institutionalreview{Not applicable.}

\dataavailability{No new data were created or analyzed in this study. Data sharing is not applicable to this article} 


\conflictsofinterest{The authors declare no conflicts of interest.}

\abbreviations{Abbreviations}{
The following abbreviations are used in this manuscript:
\\

\noindent 
\begin{tabular}{@{}ll}
HOT & Hot Optical Tube\\
LG & Laguerre-Gauss\\
BEC & Bose-Einstein Condensate\\
GPE & Gross-Pitaevskii Equation \\
\end{tabular}
}


\begin{adjustwidth}{-\extralength}{0cm}

\reftitle{References}




\isAPAandChicago{}

%


\PublishersNote{}
\end{adjustwidth}
\bibliography{References.bib}
\end{document}